\documentstyle[12pt]{article}
\begin{document}
\thispagestyle{empty}

\newcommand{\be}{\begin{equation}}
\newcommand{\ee}{\end{equation}}
\newcommand{\sect}[1]{\setcounter{equation}{0}\section{#1}}
\renewcommand{\theequation}{\thesection.\arabic{equation}}
\newcommand{\vs}[1]{\rule[- #1 mm]{0mm}{#1 mm}}
\newcommand{\hs}[1]{\hspace{#1mm}}
\newcommand{\mb}[1]{\hs{5}\mbox{#1}\hs{5}}
\newcommand{\bea}{\begin{eqnarray}}
\newcommand{\ena}{\end{eqnarray}}

\newcommand{\wt}[1]{\widetilde{#1}}
\newcommand{\und}[1]{\underline{#1}}
\newcommand{\ov}[1]{\overline{#1}}
\newcommand{\sm}[2]{\frac{\mbox{\footnotesize #1}\vs{-2}}
		   {\vs{-2}\mbox{\footnotesize #2}}}
\newcommand{\prt}{\partial}
\newcommand{\eps}{\epsilon}

\newcommand{\R}{\mbox{\rule{0.2mm}{2.8mm}\hspace{-1.5mm} R}}
\newcommand{\Z}{Z\hspace{-2mm}Z}

\newcommand{\cd}{{\cal D}}
\newcommand{\cg}{{\cal G}}
\newcommand{\ck}{{\cal K}}
\newcommand{\cw}{{\cal W}}

\newcommand{\vj}{\vec{J}}
\newcommand{\vl}{\vec{\lambda}}
\newcommand{\vz}{\vec{\sigma}}
\newcommand{\vt}{\vec{\tau}}
\newcommand{\vw}{\vec{W}}
\newcommand{\poiss}{\stackrel{\otimes}{,}}


\newcommand{\NP}[1]{Nucl.\ Phys.\ {\bf #1}}
\newcommand{\PL}[1]{Phys.\ Lett.\ {\bf #1}}
\newcommand{\NC}[1]{Nuovo Cimento {\bf #1}}
\newcommand{\CMP}[1]{Comm.\ Math.\ Phys.\ {\bf #1}}
\newcommand{\PR}[1]{Phys.\ Rev.\ {\bf #1}}
\newcommand{\PRL}[1]{Phys.\ Rev.\ Lett.\ {\bf #1}}
\newcommand{\MPL}[1]{Mod.\ Phys.\ Lett.\ {\bf #1}}
\newcommand{\BLMS}[1]{Bull.\ London Math.\ Soc.\ {\bf #1}}
\newcommand{\IJMP}[1]{Int.\ Jour.\ of\ Mod.\ Phys.\ {\bf #1}}
\newcommand{\JMP}[1]{Jour.\ of\ Math.\ Phys.\ {\bf #1}}
\newcommand{\LMP}[1]{Lett.\ in\ Math.\ Phys.\ {\bf #1}}

\renewcommand{\thefootnote}{\fnsymbol{footnote}}

\newpage
\setcounter{page}{0}
\pagestyle{empty}

\vs{30}

\begin{center}

{\LARGE {\bf $N=1,2$ Super-NLS Hierarchies}}\\
[.25cm]

{\LARGE {\bf as Super-KP Coset Reductions.}}\\[1cm]

\vs{8}

{\large {F. Toppan}}\\
\quad \\
{\em Laboratoire de Physique Th\'eorique ENSLAPP
\footnote{URA 14-36 du CNRS, associ\'ee \`a l'Ecole Normale
Sup\'erieure de
Lyon, et au Laboratoire d'Annecy-le-Vieux de Physique des Particules
(IN2P3-CNRS).},}\\
{\em ENS Lyon, 46 all\'ee d'Italie,} \\
{\em F-69364 Lyon Cedex 07, France.}\\
{E-mail: ftoppan@enslapp-ens-lyon.fr}
\end{center}
\vs{8}

\centerline{ {\bf Abstract}}

\indent

We define consistent finite-superfields reductions of the $N=1,2$
super-KP hierarchies via the coset approach we already developped for
reducing the bosonic KP-hierarchy (generating e.g. the NLS hierarchy
from the $sl(2)/U(1)-{\cal KM}$ coset). We work in a manifestly
supersymmetric
framework and illustrate our method by treating explicitly the
$N=1,2$
super-NLS hierarchies.
W.r.t. the bosonic case the ordinary covariant derivative
is now replaced  by a spinorial one
containing a spin ${\textstyle {1\over 2}}$ superfield.
Each coset reduction is associated to a rational super-$\cw$ algebra
encoding a non-linear super-$\cw_\infty$ algebra structure.
In the $N=2$ case two conjugate sets of superLax operators,
equations of motion and infinite hamiltonians in involution are
derived.
Modified hierarchies are obtained from the original ones via
free-fields mappings (just as a m-NLS equation arises by representing
the $sl(2)-{\cal KM}$ algebra through the classical Wakimoto
free-fields).
\vfill
\rightline{{\small E}N{\large S}{\Large L}{\large A}P{\small
P}-L-467/94}
\rightline{April 1994}

\newpage
\pagestyle{plain}
\renewcommand{\thefootnote}{\arabic{footnote}}
\setcounter{footnote}{0}

\sect{Introduction}

\indent

The hierarchy of integrable equations leading to a solitonic
behaviour has been a widely studied subject since the fundamental
work of Gardner, Green, Kruskal and Miura \cite{GGKM} concerning the
KdV equation. In more recent time it has
particularly deserved physicists' attention expecially in connection
with matrix models, which are a sort of effective description for
two-dimensional gravity.
In such a context the partition functions of matrix models satisfy
the
so-called Virasoro-$\cw$ constraints and can be expressed in terms of
the $\tau$-functions of the hierarchies of classical integrable
equations (for a review on that, see e.g. \cite{mar} and references
therein).\par
Looking at a classification of all possible hierarchies is therefore
a very attracting problem for both mathematical and physical
reasons.
Working in the formalism of pseudo-differential operators (PDO) such
a problem can be formalized as follows: determining
all possible algebraic constraints, consistent with the KP flows, on
the infinite fields entering the KP operator (reduction procedure).
Apart from the well-known solutions of Drinfeld-Sokolov type
\cite{DS}, which can be expressed in terms of purely differential Lax
operators, in the literature other solutions, called
multi-fields KP reductions, have been obtained
\cite{{ara},{bonora},{bonora2}}. Their basic features can be stated
as follows: in the dispersionless limit they give rise
to a Lax operator fractional in the momentum $p$. Moreover, the
algebra of Virasoro-$\cw$ constraints turns out to be a $\cw_\infty$
algebra.\par
Inspired by the works \cite{yuwu} (see also \cite{bakas}), in a
previous paper
\cite{toppan} we have shown how such reductions can be derived via a
coset construction, involving a factorization of a Kac-Moody
subalgebra out of a given Kac-Moody
or polynomial $\cw$ algebra. In our framework we have immediately at
disposal a Poisson-brackets structure providing a (multi)hamiltonian
dynamics. Furthermore, the non-linear $\cw_\infty$ algebra can be
compactely interpreted as
a finite rational $\cw$ algebra (to our knowledge, the notion of
rational $\cw $ algebra has been first introduced in \cite{feher}; in
\cite{DFSTR} it has been shown that rational $\cw$ algebras appear in
the somehow different context of coset construction; a detailed
analysis of classical rational $\cw$ algebras and their quantum
deformations has been given in \cite{feher2}).\par
Even if we have not yet attempted to give a general formal proof the
examples worked out so far strongly suggest that to each coset is
associated a corresponding KP reduction; there exists indeed a
well-defined procedure telling how to associate to a given coset a
possible KP reduction. In the absence of a general theorem the
consistency of the derived reduced operator with the
KP flows should be explicitly checked, leading to lenghty but
straightforward computations. No counterexample has been found so
far.\par
A point should be made clear: in our framework we do not need to
introduce
Dirac's brackets since we do not impose hamiltonian reductions; due
to that we are able to derive modified hierarchies
via free-field mappings provided by (the classical version of) the
Wakimoto representation \cite{waki} and their generalizations.\par
In this paper we address the problem of extending the previous
bosonic construction to the $N=1,2$ supersymmetric cases, leading to
consistent coset
reductions of the super-KP hierarchy.
The fundamental reference we will follow concerning the definition of
the KP hierarchy for odd-graded derivative is \cite{manin}.
Supersymmetric integrable hierarchies have a vast literature, see
among others e.g. \cite{others}
and
for the $N=2$ case \cite{N2}.\par
We will work in a manifestly supersymmetric formalism and illustrate
our procedure by explicitly showing
the coset derivation of the $N=1,2$ super-NLS equations. Due to the
above considerations our framework can be straightforwardly applied
to derive more complicated coset theories. \par
The basic difference with respect to the bosonic case lies in the
fact that now
the subalgebra we will factor out is generated by spin
${\textstyle{1\over 2}}$
supercurrents which enter a spinorial, fermionic, covariant
derivative.\par
The supercurrents algebra generating the $N=1$ super-NLS theory
involve two oppositely charged fermionic superfields of spin
${{\textstyle{1\over 2}}}$.
The coset construction can be performed as in the bosonic case
leading
to a non-linear super-$\cw$ algebra involving an infinite number of
primary superfields, one for each integral or half integral value of
the spin $s\geq 1$.
Such superalgebra can be regarded as a rational super-$\cw$ algebra.
It guarantees the existence of a consistent reduction of the super-KP
hierarchy, which in its turn implies the integrability of the
super-NLS equation.
The totally new feature with respect to the bosonic case, i.e. that
the subsector of
fermionic superfields only appears in the reduced super-KP operator,
will be fully discussed.\par
The first two fermionic superfields in the coset super-algebra are
the first two
(fermionic) hamiltonian densities of the super-NLS equation. Two
compatible super-Poisson brackets are derived as in the bosonic case.
A superWakimoto representation for the supercurrents algebra enables
us to introduce the
associated modified super-NLS hierarchy.\par
Our results concerning the $N=1$ super-NLS equation should be
compared with that of \cite{das} (and \cite{roe}). The equation we
derive coincide with the one
analyzed in \cite{das}. While the coefficients in \cite{das} were
suitably chosen in order to provide integrability, in our case they
are automatically furnished by the coset construction. We remark here
that the Lax operator given in
\cite{das} is of matricial type, while our super-KP reduced operator
is of scalar type. More comments on the connection of the two
approaches will be given in the text.\par
For what concerns the $N=2$ case we will make use of the formalism,
already
developped in \cite{IT} for Toda theories, based on chiral and
antichiral superfields. They are equivalent to $N=1$ superfields,
which allows us to reduce the $N=2$ case to the previous one. Any
object in the $N=2$ theory (namely superfields,
covariant derivatives, hamiltonians, Lax operators) has its chirally
conjugated counterpart.\par
The scheme of the paper is the following: in the next two sections
the bosonic construction is reviewed in detail and the basic
structures which are used
also in the super-case are discussed. We would like to point out that
some of the results here presented are new. Next, the $N=1$ formalism
is introduced and the definition of super-algebra cosets is given.
The $N=1$ super-NLS hierarchy
is analyzed. The last two sections are devoted to introduce the
formalism and extend the results to the $N=2$ case.

\sect{The coset reduction of the bosonic KP hierarchy.}

\indent

In this section we summarize the basic results of \cite{toppan}
concerning the coset reduction of the bosonic KP hierarchy.\par
Let us state the problem first: the  KP hierarchy (we follow
\cite{dickey}
and the conventions there introduced)
is defined through the pseudodifferential Lax operator
\bea
L &=& \partial + \sum_{i=0}^{\infty} U_i\partial^{-i}
\label{kp}
\ena
where the $U_i$ are an infinite set of fields depending on the
spatial coordinate $x$ and the time parameters $t_k$. Let us denote
as ${L^k}_+$ the purely differential part of the $k$-th power of the
$L$ operator; an infinite set of differential equations, or flows,
for the fields $U_i$ is introduced via the equations
\bea
{\prt L\over \prt t_k}& = &[ {L^k}_+,L]
\label{flows}
\ena
The quantities
\bea
F_k  &=&  <L^k>
\label{first}
\ena
are first integrals of motion for the flows (\ref{flows}). Here the
symbol
$<A>$ denotes the integral of the residue ($<A>=\int dw a_{-1} (w)$)
for the
generic pseudodifferential operator
$A = ...+ a_{-1} \prt^{-1} +... $.\par
An infinite set of compatible Poisson brackets structures can be
introduced, leading to a (multi)-hamiltonian structure for the flows
(\ref{flows}). The first integrals of motion  are hamiltonians in
involution with respect to
all Poisson brackets.  \par
The flows (\ref{flows}) involve an infinite set of fields. The
reduction procedure
of the KP hierarchy consists in introducing algebraic constraints on
such fields, so that only a finite number of them would be
independent.
Such constraints must be compatible with the flows (\ref{flows}). As
a final result one gets a hierarchy of integrable differential
equations involving
a finite number of fields only.\par
The canonical way to perform a reduction consists in imposing the
constraint
\bea
L^n ={L^n}_+
\label{kdvred}
\ena
which tells that the $n$-th power of $L$ is a purely differential
operator, for a given positive integer $n=2,3,...$ . Such reductions
lead to generalized KdV hierarchies: for $n=2$ one gets the KdV
equation, for $n=3$ the Boussinesq one and so on. The hamiltonian
structure for such reduced hierarchies is induced from the
hamiltonian structure of the original unreduced KP. These hierarchies
are the ones originally described by Drinfeld-Sokolov \cite{DS}.
Under the Poisson brackets structure the fields entering $L^n$
satisfy a classical
finite non-linear $\cw$ algebra (of polynomial type).\par
In the limit of dispersionless Lax equation (which is taken by
assuming the fields not depending on the spatial coordinate $x$) and
in the Fourier-transformed basis (the operator $\partial \equiv p$,
$p$ the momentum) the Lax operator $L^n$ is
just given by a polynomial in $p$ of order $n$.\par
The set of reductions given by the constraint (\ref{kdvred}) does not
exhaust the set of all possible reductions compatible with the flows
of KP.
Indeed in the literature other consistent reductions have been
discussed
(see e.g. \cite{ara},\cite{bonora}). They are called multi-fields
reductions
of KP. In the language of \cite{bonora2} they are labelled by two
positive integers $p,q$ and called generalized ($p,q$) KdV
hierarchies. For this new class of reductions there exists no integer
$n$ such that the constraint (\ref{kdvred}) holds. As a basic feature
of this new class, a non-linear $W_\infty$
algebra is associated to each reduction, instead of just the
polynomial $\cw$ algebra associated to the standard Drinfeld-Sokolov
reductions.\par
In \cite{toppan} we have shown, working out explicitly some examples,
that this new set of reductions can be derived from factoring a
Kac-Moody subalgebra
out of a given Kac-Moody or polynomial $\cw$ algebra (coset
construction).
Furthermore, the structure of non-linear $\cw_\infty$ algebra
associated to such a coset is encoded in an underlining structure of
finite rational $\cw$ algebra
(since the notion of rational $\cw$ algebra has been fully explained
in \cite{{DFSTR},{toppan}}, we will not discuss it here).
Even if we do not dispose of a formal proof telling that any coset
factorization
determines its corresponding KP reduction, we believe this statement
to be true.
Indeed, for any example of coset worked out so far we were able to
find its associated KP-reduced hierarchy.\par
Before going ahead, let us constraint $U_0\equiv 0$ in (\ref{flows})
and let us discuss the first two flows for $k=1,2$. We get
respectively
\bea
 {\partial\over\partial t_1 } U_j &=& U_j' \nonumber\\
{\partial\over\partial t_2 } U_j  &=& U_j '' + 2 U_{j+1} ' - 2
 \sum_{r=1}^{j-1} (-1)^r\left( \begin{array}{c} j-1\\ r
 \end{array}\right)
U_{j-r} \prt^r U_1
\label{eqmo}
\ena
(from now on we use the standard convention of denoting
the spatial derivative with a prime and the time derivative with a
dot if no confusion concerning the flow arises)
for any $j=1,2,...$ .\par
The first flow is trivial, while the second one provides a set of
non-linear
equations. \par
For later convenience (and in order to derive the KP reduction we are
going
to discuss from an underlininig coset algebra which provides the
hamiltonian structure)
let us
introduce at this point a covariant derivative $\cal D$ (whose
precise definition will be given later), acting on covariant fields
with definite charge.
An important point is that the covariant derivative satisfies the
same rules, in particular the Leibniz rule, as the ordinary
derivative and coincides with the latter one when acting upon
chargeless fields.\footnote{the following
discussion will be limited to covariant derivatives defined for an
abelian $U(1)-{\cal KM}$ algebra, even if the non-abelian case can be
considered on the same foot as well.} At a formal level, the formulas
giving the action of covariant derivatives on covariant fields look
the same as those involving ordinary derivatives.
An example of that is the following important commutation rule
\bea
{\cal D}^{-k} f&=& f{\cal D}^{-k} +\sum_{r=1}^\infty(-1)^r
\left( \begin{array}{c} k+r-1\\ r \end{array}\right)
f^{(r)}
 {\cal D}^{-k-r}
\label{comrul}
\ena
(here $f^{(r)} \equiv {\cal D}^r f $ and $k$ is a positive
integer).\par
A consistent reduced version of the KP hierarchy can be expressed as
the Lax
operator
\bea
L &=& {\cal D} + J_- {\cal D}^{-1} J_+\equiv \partial +J_-\cdot {\cal
D}^{-1}J_+\label{nls}
\ena
Let us introduce the composite fields $V_n = J_- \cdot {\cal D}^n
J_+$. The reduction (\ref{nls}) implies the identification
\bea
U_{n} &=& (-1)^{n-1}V_{n-1}, \quad\quad n=1,2,...\label{subst}
\ena
where the $U_n$'s are the fields appearing in (\ref{kp}). It can be
easily checked
that the above position is indeed a reduction, namely
that is consistent with the flows (\ref{flows}); this statement is
proved as follows: at first one should notice that, due to the
properties of the covariant derivative, an
algebraic relation holds
\bea
 V_{p+1} \cdot V_{0} &=& V_{0} \cdot \partial V_{p} + (V_1-
\partial V_{0}) V_{p}
\label{alg}
\ena
which allows to algebraically express the fields $V_p$, for $p\geq
2$, in terms of the fundamental ones $V_0 $ and $V_1$.  Due to
standard properties of the Newton binomial, the equations for $j > 2$
in the flows
(\ref{eqmo}, $b$) are compatible with the algebraic relation
(\ref{alg}) after
taking into account the substitutions (\ref{subst}). \par
So far we have discussed the reduction of the KP hierarchy at a
purely algebraic
level, without mentioning any hamiltonian structure. Up to now the
introduction of a covariant derivative was not effective since,
as we have already remarked, covariant and ordinary derivatives play
the same role
if only algebra is concerned. The introduction of a covariant
derivative is at least a very convenient tool to make contact with
the hamiltonian dynamics and it proves to be crucial for regarding
the (\ref{nls}) reduction as a coset
construction.\par
Let us assume the fields $J_\pm (x), J_0(x)$ to satisfy the $sl(2)$
Kac-Moody algebra
\bea
\{J_+(z),J_-(w)\} &=&  \partial_w\delta(z-w) - 2 J_0 (w) \delta(z-w)
\equiv
\cd (w)\delta(z-w) \nonumber \\
 \{J_0(z), J_\pm (w)\} &=& \pm J_\pm (w) \delta (z-w)
 \nonumber \\
 \{J_0 (z),J_0(w)\} &=& {-\textstyle{1\over
 2}}\partial_w\delta(z-w)\nonumber\\
\{J_\pm(z),J_\pm(w)\} &=& 0
\label{kmalg}
\ena
the covariant derivative $\cal D$ is defined acting on covariant
fields $\Phi_q$ of definite charge $q$ as
\bea
{\cal D} &=& (\partial +2q J_0)\Phi_q
\ena
The property of covariance for the field $\Phi_q$ being defined
through the relation
\bea
\{ J_0 (z), \Phi_q (w) \} &=& q \Phi_q(w)\delta (z-w)
\ena
As its name suggests, the covariant derivative maps covariant fields
of charge $q$ into new covariant fields having the same charge.
In particular $J_\pm $ are covariant fields with respect to $J_0$ and
have charge $\pm 1$ respectively, so that
\bea
{\cal D} J_\pm&= & \partial J_\pm \pm 2 J_0 \cdot J_\pm
\ena
The algebraic relations (\ref{kmalg}) of the $sl(2)$-Kac-Moody can be
seen as a
first Poisson bracket structure (denoted as $ \{\cdot, \cdot \}_1$)
for the reduced (\ref{nls}) hierarchy. It is a trivial check indeed
to show that the
first two integrals of motion $F_{1,2}$ (\ref{first}) are
proportional to $H_{1,2}$:
\bea
H_1&=&\int (J_-\cdot J_+)\nonumber\\
H_2&=& -\int (J_-\cdot {\cal D}J_+)
\label{hami}
\ena
which are hamiltonians in involution with respect to the
(\ref{kmalg}) Poisson brackets; $H_{1,2}$ reproduce respectively the
first and the second flow of (\ref{eqmo}) under the substitution
(\ref{subst}):
\bea
 {\partial\over\partial t_1 } V_n &=& \{ H_1, V_n\} = V_n '
 \nonumber\\
{\partial\over\partial t_2 } V_n  &=& \{ H_2, V_n\} =
V_n '' -2V_{n+1} ' -2
\sum_{r=1}^n \left( \begin{array}{c} n\\ r \end{array}\right)
V_{n-r} \prt^rV_0
\label{eqmo2}
\ena
\par
for $n=0,1,...$ .\par
Our framework allows us to accomodate a second compatible Poisson
brackets structure which is given by
\bea
\{ J_-(z), J_- (w)\}_2&=& 0\nonumber\\
\{ J_+ (z), J_+ (w) \}_2 &=& -\delta (z-w) (J_+)^2 (w)\nonumber\\
\{ J_+ (z), J_- (w) \}_2 &=& {{\cal D}_w}^2 \delta (z-w) +\delta
(z-w) J_+(w)J_-(w)
\ena
To understand the above relations, one should notice that they are
obtained
from the corresponding relations for the first Poisson brackets
structure (\ref{kmalg}) after taking into account the substitutions
\bea
J_- &\mapsto& J_-\nonumber \\
J_+ &\mapsto &-{\cal D}J_+
\ena
The compatibility of first and second Poisson brackets simply means
the following equality being satisfied
\bea
{\dot f} &=& \{ H_1, f\}_2 = \{ H_2, f\}_1
\ena
\par
The composite fields $V_n$ entering (\ref{kmalg}) are by construction
chargeless, i.e. they have vanishing Poisson brackets with respect to
$J_0$
\bea
\{ J_0 (z), V_n (w) \} &=& 0
\label{comm}
\ena
They constitute a linearly independent basis for the composite
chargeless bilinear fields (bilinear invariants); namely any such
field can be obtained as a linear combination of the $V_n$ fields and
ordinary derivatives acting on them. Under the first Poisson brackets
structure the fields $V_n$ form a closed
non-linear algebra. The only finite subset which is closed with
respect to this algebra is given by $V_0$ itself: as soon as every
other field is added, one needs the whole infinite set of fields to
close the algebra. These bilinear invariants therefore provide the
reduction (\ref{nls}) with the structure of a non-linear $W_\infty$
algebra.  Since however the $V_n$ fields, even if linearly
independent,
are not algebraically independent due to relations like (\ref{alg}),
the
non-linear $\cw_\infty$ algebra structure can be regarded as
encoded in the more compact structure of finite rational $\cw$
algebra.
For more details and for the explicit expression of such rational
algebra see \cite{toppan}.
 \par
The fact that the fields $V_n$ have vanishing Poisson brackets with
respect to the $U(1)-{\cal KM}$ subalgebra of the Kac-Moody $sl(2)$
means that we have found the explicit link between our KP-reduction
(\ref{nls}) and the coset
factorization.\par
In \cite{toppan} another such reduction was considered in full
detail; it was associated to the Lax operator
\bea
{\tilde L}& =& {\cal D}^2 + T + W_-\cdot {\cal D}^{-1} W_+
\label{op}
\ena
Such operator has not the form of a KP operator, but it is however
possible to introduce the uniquely defined ``square root" $ {\tilde
L}^{{1\over 2}}$ of ${\tilde L}= {\tilde L}^{{1\over 2}} \cdot
{\tilde L}^{{1\over 2}}$ which is of KP-type (${\tilde L}^{{1\over
2}}= {\cal D} +... $). The fields $T, W_\pm$ entering (\ref{op}), are
respectively a
chargeless stress-energy tensor and two (opposite charged) bosonic
spin ${\textstyle{3\over 2}}$ fields;
the charge being defined with respect to an $U(1)-{\cal KM} $
current $J$ entering the covariant derivative. The fields $J, T,
W_\pm$ form a closed algebra which is nothing else that the
non-linear Polyakov-Bershadski
$\cw$ algebra \cite{polya}. It plays
the same role of first Poisson brackets structure leading to a
hamiltonian dynamics for
the flow associated to the (\ref{op}) Lax operator, just like the
$sl(2)-{\cal KM}$ algebra in the previous case. The same steps done
before can be repeated in this case too.\par
In general, starting from a given coset algebra, it is quite an easy
Ansatz to find out the form of the reduced KP Lax operator; the
following
steps should be performed: at first the Kac-Moody currents of the
factorized subalgebra should be accomodated into a single covariant
derivative, then with the help of dimensional considerations one
should identify the $U_n$ fields of (\ref{kp}) with invariants
constructed out of covariant fields, the original ones in the algebra
as well as the covariant derivatives applied on them. The only
difficulty left consists in explicitly checking the consistency of
such reduction
with the KP flow, as well as its link with the hamiltonian dynamics
provided by the algebra itself.\par
We close this section with a remark: in the limit of dispersionless
Lax equation,
and taking into account that $J_0$ is a constant ($\equiv \alpha$)
with respect to any flow
due to the relations (\ref{comm}), the reduced operators (\ref{nls})
and (\ref{op}) are respectively given by
\bea
L&\rightarrow& p + {\lambda \over p+ \alpha} \nonumber\\
{\tilde L}& \rightarrow & p^2 + t +{\lambda \over p+\alpha}
\ena
with $\alpha ,\lambda$ and $ t $ constants.\par
It is remarkable that the reductions associated to rational $\cw$
algebras lead,
in the dispersionless limit, to Lax operators fractional in $p$
(this is always true in any case of coset construction), while the
Drinfeld-Sokolov reductions associated to polynomial $\cw$ algebras
lead to Lax operators polynomial in $p$.

\section{From NLS to a modified NLS hierarchy via Wakimoto
representation of the $sl(2)-{\cal KM}$
algebra.}

\indent

In this section we study more closely the hierarchy associated
to the reduced KP operator (\ref{nls}). We show that it coincides
with the two-components formulation of the NLS hierarchy. In terms of
the second hamiltonian $H_2 = -\int (J_- {\cal D } J_+)$ we get
indeed the following equations
\bea
{\dot {J_\pm}}&= &  \{ J_\pm , H_2\}_1=\pm {\cal D}^2 J_\pm \pm 2
(J_+ J_-)J_\pm
\label{nls2}\ena
This is the coupled system associated to the NLS equation. Due to the
results
mentioned
in the previous section it is consistent to set $J_0\equiv 0$, which
further implies ${\cal D}^2 J_\pm = {J_\pm}'' $.
Next, the standard NLS equation is recovered by letting the time
being imaginary. Such ``Wick rotation" allows making the
identification
\bea
{J_-}^\star &=& J_+ = u
\ena
We obtain finally
\bea
i {\dot u} &=& u'' + 2 u|u|^2
\end{eqnarray}
which is the NLS equation in its standard form \cite{fadtak}.\par
At this point we should recall that equivalent integrable equations
can arise in two different ways: either because they are associated
to different hamiltonians
belonging to the same hierarchy of hamiltonians in involution, or
because there exists a mapping between them. This is the case
concerning the relation between
KdV and m-KdV equations, the latter being the equation involving the
free field
$\varphi$, which is related via Miura transformation to the $v$ field
satisfying the KdV equation; for the KdV Lax operator this reads as
follows
\bea
\partial^2 + v& =&(\partial-\varphi)(\partial + \varphi ) =
\partial^2 +\varphi ' - \varphi^2
\ena
Generalizations of this construction hold for any hierarchy of
Drinfeld-Sokolov type.\par
The framework we developped in the previous section is particularly
useful
for describing the analogue free-fields mappings in the case of coset
reductions.
There exists indeed a standard free field representation of the
$sl(2)-{\cal KM}$
algebra which is given by the (classical) Wakimoto representation
\cite{waki}. It is realized
in terms of the weight $1$ field ${\nu} $\footnote{ in the standard
notation for the quantum Wakimoto representation $\nu \equiv \partial
\phi$, $\phi$ is the fundamental field satisfying the OPE $\phi
(z)\phi(w)\sim log (z-w)$.}
and the bosonic $\beta-\gamma$ system
of weight $(1,0)$, satisfying the algebra
\bea
\{\beta (z) , \gamma (w) \} &=& -\{ \gamma (z) , \beta (w) \} =
\delta (z-w)\nonumber\\
\{\nu (z), \nu (w) \} &=& \partial_w \delta (z-w)
\ena
(any other Poisson bracket is vanishing).\par
The $sl(2)-{\cal KM}$ algebra given in (\ref{alg}) is reproduced
through
the identifications
\bea
J_+&=&\beta\nonumber\\
J_0 &=& -\beta \gamma + {i\over {\sqrt 2}}\nu \nonumber\\
J_- &=& \beta \gamma^2 -i{\sqrt 2} \gamma \nu +\partial\nu
\ena
Representing the hamiltonian $H_2 $ in terms of the Wakimoto fields,
one can derive the coupled system
\bea
{\dot {\beta }}&= & \{\beta, H_2\}_1 = \beta '' + 2\beta^2\gamma '
-2\beta^3\gamma^2\nonumber\\
{\dot {\gamma}} &=& \{\gamma , H_2\}_1=\gamma '' -2\gamma^2\beta '
-2\gamma^3\beta^2
\label{mnls}
\ena
(we used here the consistent constraint $J_0 = 0$ to get rid of the
field $\nu$ in the above equations). The fields $\beta , \gamma$
enter in a symmetric way in the above system and we can forget about
the different weights between them
we originally introduced to define the Wakimoto representation. If we
let indeed the spatial coordinate being imaginary ($\partial_x
\mapsto i\partial_x $), it is consistent to set
\bea
\gamma^\star&= &\beta = \lambda
\ena
so that the final result is
\bea
{\dot {\lambda}}& =& -\lambda '' + 2\lambda
(i\lambda\partial\lambda^\star -
|\lambda|^4 )
\label{redmnls}
\ena
\par
It is a remarkable fact that the modified NLS system (\ref{mnls})
should be regarded
as some sort of dual version of the original NLS system (\ref{nls2}).
In the latter case the reduction to the single component NLS equation
is done by assuming the time being imaginary, while in the m-NLS case
this is provided by assuming the space being imaginary. \par
The construction here discussed can be trivially extended to any
coset arising from generic Kac-Moody algebra. The free-fields
analogue of the Wakimoto representation is in this case provided by
(the classical version of) the results of ref. \cite{GMMOS}.\par
Let us finally stress the point that in our approach to the KP-coset
reduction
the connection with the free-fields representation is particularly
explicit,
since we did not need to introduce any Dirac brackets arising from
the constraint
$J_0\equiv 0$: in our framework all computations are performed using
the original Poisson brackets structure.

\section{The coset derivation of the $N=1$ super-NLS equation.}

\indent

In this section we will set up a manifestly supersymmetric framework
to derive via
coset construction $N=1$ supersymmetric integrable hierarchies. There
are two basic motivations for doing that. The first concerns of
course the construction of
superintegrable hierarchies, which are interesting by their own, and
have been widely studied in the literature (see e.g.
\cite{{others},{N2}}). The second motivation lies in better
understanding the coset construction itself. Before any attempt of
classifying
the cosets and before giving general formal proofs of their link with
the hierarchies, it is interesting to investigate  how they look in
the case of superalgebras.\par
It should be kept in mind that even if our discussion will concern
the super-NLS hierarchy only, in no respect this example is crucial.
The same approach here discussed
can be straightforwardly applied to derive other supersymmetric coset
hierarchies. It is enough for that to apply the machinery here
developped to any given coset algebra.
The advantage of discussing the super-NLS case lies in its technical
simplicity. \par
The super-NLS case is however not an academical exercise and it is
interesting
to compare our results with that of \cite{{roe},{das}}. In \cite{roe}
two distinct supersymmetrizations, one of these involving a free
parameter, of the NLS equation have been proposed. It is stated that
both
lead to an integrable hierarchy. In \cite{das} manifestly
supersymmetric NLS equations have been investigated. It has been
shown that applying on such equations conventional tests of
integrability only the supersymmetric system without any free
parameter is selected. Moreover there exists a discrepancy in the
coefficients with respect to \cite{roe}. The coset construction we
are going to discuss will automatically provide the super-NLS
integrable system
of ref. \cite{das} with the same coefficients (therefore supporting
the statement of \cite{das} that a misprint occurs in \cite{roe}).
Our coset construction implies
that associated to such system there exists a non-linear
super-$\cw_\infty$ algebra
involving an infinite series of primary bosonic (of integral
dimension $h=1,2, ... $) and fermionic (of half-integral dimension $
h={\textstyle{3\over 2}}, {\textstyle{5\over 2}}, ...$) $N=1$
superfields. Such super-$\cw_\infty$
algebra can be regarded as a rational super-$\cw$ algebra. The
existence of this
non-linear super-$\cw_\infty$ algebra is already an indication of the
integrability properties of our super-NLS system. This statement is
made precise by associating to
the coset a
consistent reduction of the super-KP hierarchy. Our Lax operator is
different from the one discussed in \cite{das}.
\par
Let us fix now our conventions concerning the superspace. We denote
with capital
letters the $N=1$ supercoordinates ($X\equiv, x, \theta$, with $x$
and $\theta
$ real, respectively bosonic and grassmann, variables). The
supersymmetric spinor
derivative is given by
\bea
D \equiv D_X &=& {\partial\over \partial\theta} +\theta
{\partial\over \partial x}
\ena
With the above definition $ {D_X}^2 ={\textstyle{\partial\over
\partial x}}$. \par
The supersymmetric delta-function $\Delta (X,Y)$ is a fermionic
object
\bea
\Delta (X,Y) &=& \delta (x-y) (\theta -\eta)
\ena
It satisfies the relations
\bea
\Delta (X,Y) &=& -\Delta (Y,X) \quad\quad\quad
D_X\Delta (X,Y) - D_Y\Delta (X,Y)
\ena
Our convention for the integration over the grassmann variable is
\bea
\int d\theta \cdot \theta &=& -1
\ena
For any given superfield $F(X)$ we get then
\bea
\int dY \Delta (X, Y )F(Y) &=& F(X)
\ena
As in the bosonic case, the (super)-line integral over a total
derivative gives a vanishing result.
The canonical dimensions $d$ are respectively given by
\bea
d(D) &=& d(\Delta) = -d(\theta) = -2d(x) ={\textstyle {1\over 2}}
\ena
The role which in the bosonic case is played by the ordinary
derivative is now
played by the spinor derivative of dimension $d={\textstyle {1\over
2}}$. It makes plausible that now covariant spinor derivatives should
be constructed in terms of
spin ${{\textstyle {1\over 2}}}$ fermionic superfields. An example of
supersymmetric rational $\cw$ algebra involving such kind of
derivatives has indeed been given in \cite{DFRS}. \par
The $N=1$ counterpart of the $U(1)-{\cal KM}$ current $J_0(z)$ should
be expressed by the fermionic superfield $\Psi_0(X)= \psi_0 (x) +
\theta J_0(x)$, satisfying the super-Poisson brackets
relation\footnote{
we recall that super-Poisson brackets are symmetric  when taken
between odd elements, antisymmetric otherwise.}
\bea
\{ \Psi_0 (X), \Psi_0 (Y) \} &=& D_Y \Delta (X,Y)
\label{zerosusyalg}
\ena
which implies, at the level of components
\bea
\{\psi_0(x),\psi_0(y)\}&=&-\delta (x-y)\nonumber\\
\{J_0(x),J_0(y)\}&=&-\partial_y\delta (x-y)
\ena
Super-covariant fields and the supercovariant derivative can now be
introduced
through
\bea
\{ \Psi_0 (X), \Phi_q (Y) \} &=& q\Delta (X,Y) \Phi_q (Y)\nonumber\\
{\cal D}\Phi_q &=& D\Phi _q + q \Psi_0 \Phi_q
\ena
$\Phi_q$ is a covariant superfield (either bosonic or fermionic).\par
We are now in the position to discuss the algebra providing the first
(super)-Poisson brackets structure for the super-NLS equation. As
suggested in \cite{das}, the component fields should be accomodated
in two fermionic spin ${\textstyle{1\over 2}}$
superfields $\Psi_\pm = \psi_\pm + \theta J_\pm $. With the above
choice one can identify the bosonic components $J_\pm$ with the
analogue fields we already
encountered in the bosonic case. The relevant algebra can therefore
be simply guessed
to be the supersymmetric analogue of the $sl(2)-{\cal KM}$ algebra,
introduced through the relations
\bea
\{ \Psi_0 (X), \Psi_\pm (Y) \} &=& \pm \Delta (X,Y) \Psi_\pm
(Y)\nonumber \\
\{ \Psi_+ (X), \Psi_- (Y) \} &=& {\cal D}_Y \Delta (X,Y) = D_Y \Delta
(X,Y)
+ \Delta (X,Y) \Psi_0(Y)
\label{susyalg}
\ena
One indeed recover the (\ref{kmalg}) algebra by setting all the
component spin
${\textstyle{1\over 2}}$ fermionic fields equal to $0$.\par
We can define, just like in the bosonic case, the composite
superfields $V_n (X)$,
where
\bea
V_n &=& \Psi_- {\cal D}^{n} \Psi_+ \quad\quad n=0,1,2,...
\label{superinv}
\ena
By construction they have vanishing Poisson brackets with respect to
$\Psi_0 $:
\bea
\{ \Psi_0 (X), V_n (Y) \} &=& 0
\ena
The superfields $V_n$ are respectively bosonic for even values of $n$
and fermionic for odd values. They play the same role as the
corresponding fields
in the purely bosonic case: they constitute a basis of linearly
independent superfields for the chargeless composite superfields. The
super Poisson brackets (\ref{zerosusyalg},\ref{susyalg}) provide such
basis of fields with the structure of non-linear super-$\cw_\infty$
algebra that will be discussed later in more detail. \par
In order to associate to the coset algebra a hamiltonian dynamics
like we did in the bosonic case we proceed as follows: we recall that
the superfields $V_n$
have positive dimensions $d(V_n) = {\textstyle { n+2\over 2}}$; then
we look for
all possible hamiltonian densities of a given dimension that one can
algebraically construct out of the superfields $V_n$ and the
covariant derivative (of dimension ${\textstyle {1\over 2}}$) acting
upon them. For any given
dimension only a finite number of such combinations are allowed.
Since now we are working in a manifestly supersymmetric framework and
the (super)-line integral is fermionic, so the hamiltonian densities
must be fermionic of half-integral dimension. The first two
possible hamiltonian densities at dimension ${\textstyle {3\over 2}}$
and ${\textstyle {5\over 2 }}$ respectively are just given by $V_1$
and $V_3$. The latter is indeed the unique, up to a total derivative,
chargeless $d ={\textstyle{5\over 2}}$ object.\par
It can easily checked now that $H_{1,2}$ given by
\bea
H_1 &=& \int dX V_1 (X) = \int dX (\Psi_- \cdot {\cal
D}\Psi_+)\nonumber\\
H_2 &=& \int dX V_3 (X) = \int dX (\Psi_-\cdot {\cal D}^3 \Psi_+)
\label{superhami}
\ena
have vanishing Poisson brackets among themselves
with respect to (\ref{zerosusyalg},\ref{susyalg}) and can therefore
been regarded as hamiltonians in involution. Two compatible flows are
defined through
\bea
{\partial \over \partial t_1 } \Psi_\pm &=& \{ H_1, \Psi_\pm \} =
{\cal D}^2 \Psi_\pm
\nonumber\\
{\partial \over \partial t_2 } \Psi_\pm &=& \{ H_2, \Psi_\pm \} =
\pm {\cal D}^4 \Psi_\pm \mp \Psi_\pm { D}( \Psi_\mp {\cal D} \Psi_\pm
)
\label{superNLS}
\ena
The latter equation is the $N=1$ supersymmetric version of the
two-components NLS.  As in the bosonic case, if we let the time $t_2$
be imaginary we can consistently set
\bea
 \Psi_+ = {\Psi_-}^\star = \Psi
\nonumber
\ena
to get the super-NLS equation
\bea
i{\dot \Psi} &=& \Psi^{(4)} -\Psi D(\Psi^\star \Psi^{(1)})
\ena
(in order to simplify the notation from now on the symbol
$A^{(n)} \equiv {\cal D}^{n} A $ will be used).  \par
Since ${\dot \Psi_0}=0$ makes consistent to set $\Psi_0 =0$, the
above equation
leads to the following system in component fields ($\Psi= \phi
+\theta q$,
 $\phi$ fermionic and $q$ bosonic):
\bea
i{\dot \phi} &=& \phi_{xx} + \phi ( \phi^\star \phi_x - q^\star q
)\nonumber \\
i{\dot q } &=& q_{xx} - (q q^\star) q + (\phi {\phi_x}^\star
-\phi_x\phi^\star) q
+(\phi \phi^\star)q_x
\ena
As already stated, this equation coincides with the integrable
super-NLS equation of ref. \cite{das}.\par
The supersymmetric character of the above equations is guaranteed by
the invariance of the hamiltonians $H_{1,2}$ under the
transformations
\bea
\delta \Psi_\pm &=& \pm\varepsilon {\cal D}\Psi_\pm
\ena
where $\varepsilon $ is a grassmann parameter.\par
The existence of a bihamiltonian structure is derived as in the
bosonic case.  The second super-Poisson brackets structure is given
by
\bea
\{ \Psi_- (X), \Psi_- (Y) \}_2 &=& 0\nonumber\\
\{\Psi_- (X), \Psi_+ (Y) \}_2 &=& \Delta^{(3)}
-\Delta^{(1)}\Psi_-\Psi_+ +\Delta {\Psi_-}^{(1)}\Psi_+\nonumber\\
\{\Psi_+ (X), \Psi_+ (Y) \}_2 &=& \Delta^{(2)} \Psi_+{\Psi_+}^{(1)}
-\Delta^{(1)}\Psi_+{\Psi_+}^{(2)} +\Delta
{\Psi_+}^{(1)}{\Psi_+}^{(2)}
\ena
(The superfields on the right hand side are evaluated in $Y$ and
$\Delta^{(n)}=
{{\cal D}_Y}^{n} \Delta (X,Y) $).\par
This second Poisson brackets structure is derived from the first one
after the substitutions
\bea
\Psi_- &\mapsto & \Psi_-
\nonumber\\
\Psi_+ &\mapsto& {\cal D}^{2} \Psi_+
\ena
are taken into account. \par
The compatibility of the two Poisson brackets structure is ensured,
like in the bosonic case, by the relation
\bea
{{d F\over d t}} &=& \{ H_1, F\}_2 = \{ H_2, F\}_1
\end{eqnarray}
Precisely like the bosonic case, the two hamiltonians $H_{1,2}$ are
the first two
of an infinite series of hamiltonians mutually in involution. This
statement will be justified later when we show how to
associate to the system (\ref{superNLS}) a reduction of the super-KP
hierarchy.\par
A comment is in order. The algebra (\ref{zerosusyalg},\ref{susyalg})
is the simplest possible
algebra realized in terms of supercurrents and allowing a Kac-Moody
coset construction. There is another very simple supercurrent
algebra, which is
realized by just coupling to the $\Psi_0$ superfield two bosonic
superfields $\Phi_\pm$ (instead of two fermionic ones) of dimension
${\textstyle{1\over 2}}$. The expression of this algebra
looks like (\ref{susyalg}) but now one has to take into account the
antisymmetric
property when exchanging $\Phi_\pm$ in the super-Poisson brackets. If
we define the charges being the super-line integral over the
supercurrents (as $H=\int dX \Psi_0$, $E_\pm = \int dX \Psi_\pm$),
then the algebra (\ref{zerosusyalg},\ref{susyalg}) generates
a global $sl(2)$ algebra for the charges, while the algebra
determined by $\Psi_0$ and the bosonic supercurrents is promoted to
the global superalgebra $osp(1|2)$ (with generators $H$, $F_\pm =
\int dX \Phi_\pm$). In \cite{das} a zero-curvature
formulation for the system (\ref{superNLS}) was found; it is based on
the
$sl(2)$ algebra. They
claimed being unable to derive an analogue formulation starting from
$osp(1|2)$.
The reason is simply because this is associated with a radically
different system, the dynamics being in this case defined for the
bosonic $\Phi_\pm $ superfields.
The fact that the dynamics differs from the fermionic case can be
immediately
seen using the following argument: an invariant composite superfield
$W_0 \cdot W_1 $ ($W_n = \Phi_-{\cal D}^n \Phi_+$) is allowed
entering in the second hamiltonian density of dimension
${\textstyle{5\over 2}}$, while the corresponding composite
superfield $V_0\cdot V_1$ is vanishing in the fermionic case for the
antisymmetry
of $\Psi_\pm$. Our coset construction can be performed for this
bosonic case as well, leading to an interesting superintegrable
system, which of course has nothing to do with the
supersymmetrization of the NLS equation, since the component bosonic
fields have spin ${\textstyle{1\over 2}}$ and not $1$. It is likely
that for such a system the zero-curvature formulation would be based
on the superalgebra $osp(1|2)$. We leave a detailed discussion of it
for a further
publication.

\section{Comments on the non-linear super-$\cw_\infty$ coset
algebra.}

\indent

Let us make some more comments here concerning the non-linear
super-$\cw_\infty$ algebra structure of the coset algebra. Its linear
generators are the superfields
$V_n$, $n$ non-negative integer, defined in (\ref{superinv}). The
superfields are
bosonic for even values of $n$, fermionic for odd values. The set
$\{V_0, V_1\}$
constitutes a finite super-algebra, given by the Poisson brackets
\bea
\{ V_0 (X), V_0(Y) \} &=& -\Delta (X,Y) (DV_0 +2V_1)(Y)\nonumber\\
\{ V_0 (X), V_1(Y) \} &=& \Delta^{(2)} (X,Y))
V_0(Y) +\Delta^{(1)}(X,Y) V_1(Y) -\Delta (X,Y) DV_1 (Y)\nonumber\\
\{ V_1 (X), V_1(Y) \} &=& -2\Delta^{(2)} (X,Y) V_1(Y) -\Delta (X,Y)
{D_Y}^2V_1(Y)
\label{supcos}
\ena
In terms of component fields it is given by two bosons of spin $1$
and $2$
respectively, and two spin ${\textstyle {3\over 2}}$ fermions. It is
the maximal
finite subalgebra of the coset superalgebra: as soon as any other
superfield is added
to $V_0,V_1$, the whole set of fields $V_n$ is needed to close the
algebra, giving to the coset the structure of a super-$\cw_\infty$
algebra. Moreover such algebra closes in non-linear way.
\par
Using the techniques developped in \cite{DFSTR} it is possible to
show the existence of an equivalent basis for expressing our
super-$\cw_\infty$ algebra,
given by the infinite set of superfields $W_h (X)$, which are primary
with
conformal dimension $h$ with respect to the stress-energy tensor
(having vanishing central charge) $T(X) \equiv W_{\textstyle{3\over
2}}(X)$.
To any integral value of $h$ ($h=1,2,..$) is associated a bosonic
primary superfield;
to any half-integral value ($h ={\textstyle{3\over 2}},
{\textstyle{5\over 2}},... $) a fermionic one. \par
The condition of being primary means that the superfields $W_h$
satisfy the relation
\bea
\{ T(X), W_h(Y) \} &=& -{h}\Delta^{(2)}(X,Y) W_h(Y) +{1\over 2}
\Delta^{(1)}(X,Y) DW_h (Y)-\Delta (X,Y)
D^2 W_h (Y)\nonumber\\
\ena
We have at the lowest orders
\bea
W_1 &=& V_0 =\Psi_- \Psi_+\nonumber \\
T &=& V_1 + {\textstyle{1\over 2 } }DV_0 = \nonumber\\
&=&
{\textstyle{1\over 2 } }
{\cal D} {\Psi_-}\cdot\Psi_++
{\textstyle{1\over 2 } }
\Psi_-\cdot {\cal D}{\Psi_+}\nonumber\\
W_2 &=& 3V_2 + DT -{\textstyle {3\over 2}} \partial V_0 =
\nonumber\\
&=& \Psi_-\cdot
{\cal D}^2{\Psi_+} +{\cal D}{\Psi_-}\cdot {\cal D} {\Psi_+}
-{\cal D}^2{\Psi_-}\cdot {\Psi_+}
\ena
We wish finally to make some comments on the rational character of
the above defined super-$\cw_\infty$ algebra: the whole set of
algebraic relations can be
expressed just in terms of closed rational super-$\cw$ algebra
involving $4$ superfields as the following reasoning shows: let us
introduce the superfields
\bea
\Lambda_p &=_{def}& {\cal D} \Psi_- \cdot {\cal D}^{(p+1)}
\Psi_+\nonumber
\ena
then
\bea
\Lambda_p &=& D V_{p+1} - V_{p+2} \nonumber
\label{lambdadef}
\ena
Due to standard properties of the covariant derivative we can write
down for the superfields $\Lambda_p$ the analogue of the relation
(\ref{alg}) of the bosonic case:
\bea
\Lambda_0 \Lambda_{p+1} &=& \Lambda_0 D\Lambda_p + (\Lambda_1 -
D\Lambda_0)\Lambda_p
\label{ratio2}
\ena
which implies that $\Lambda_p$ are rational functions of
$\Lambda_{0,1}$, which
in their turns are determined by $V_i$, $i=0,1,2,3$. \\
Inverting the relation (\ref{lambdadef}) we can express any higher
field $V_{p+1}$ in terms of $V_p$, $\Lambda_{p-1}$. As a consequence
of this we have
the (rational) closure of the superalgebra on the superfields
$V_0,V_1, V_2, V_3$. \par
We remark that now is not possible, like in the bosonic case, to
determine higher
order superfields $V_p$ from the formula (\ref{ratio2}) by simply
inserting $V_0$, $V_p$ in place of $\Lambda_0$, $\Lambda_p$: this is
due to the fact that
any product $V_0\cdot V_{p+1} $ identically vanishes since it is
proportional to a squared  fermion (${\Psi_-}^2=0$).  That is the
reason why four superfields are
necessary to produce a finite rational algebra and not just two as
one would
have nively expected.

\section{The $N=1$ superWakimoto representation and the modified
super-NLS equation.}

\indent

In this very short section we will repeat the construction of section
3, furnishing the $N=1$ super-Wakimoto representation of the
(\ref{zerosusyalg},\ref{susyalg}) algebra
and associating to the super-NLS equation (\ref{superNLS}) its
modified version.\par
The classical super-Wakimoto representation is realized in terms of
three
free superfields, denoted as $B, C, N$:
\bea
B(X) &=& b(x) + \theta \beta (x)\nonumber\\
C (X) &=& \gamma (x) + \theta c (x)\nonumber\\
N(X)& =& \mu (x) + \theta \nu (x)
\ena
$B, N $ are assumed to be fermionic of dimension ${\textstyle {1\over
2}}$,
while $C$ is assumed to be a $0$-dimensional bosonic superfield
coupled to $B$.
At the level of components we have in particular the already
encountered
bosonic $\beta - \gamma$ system of weight $(1,0)$, plus now a
fermionic $b-c$
system of weight $({\textstyle {1\over 2}},{\textstyle {1\over
2}})$.\par
The free superfields super-Poisson brackets are given by
\bea
\{ B (X), C (Y) \} &=&
\{ C(X), B(Y) \} =\Delta (X,Y)\nonumber \\
\{ N (X), N (Y) \} &=& D_Y \Delta (X,Y)
\ena
The (\ref{zerosusyalg},\ref{susyalg}) superalgebra is reproduced in
terms of the superfields $B,C,N$
through the identifications
\bea
\Psi_+ &=& B\nonumber\\
\Psi_0 &=& - B C +  N\nonumber
\\
\Psi _- &=& -{1\over 2} B C^2 +  C N - DC
\label{superwak}
\ena
Representing $H_2$ in (\ref{superhami}) via the above system we get
an evolution
equation for $B,C$. As in the bosonic case the $N$ superfield can be
expressed
through $B,C$ by setting $\Psi_0=0$. Finally, by letting the space
being imaginary it is consistent to further set
\bea
({\cal D }B)^\star &=& C
\ena
which implies
\bea
\beta (x) &=& \gamma (x); \quad\quad b' (x) = c(x)
\ena
At the end we arrive at the supersymmetric generalization of eq.
(\ref{redmnls}), which is given by
\bea
{\dot B} &=& - D^4B + B ({D}(C^\star D C ) - {\textstyle{1\over 2 }
}|C|^4  )
\ena

\section{Integrable properties of the $N=1$ super-NLS equation: the
super-KP reduction.}

\indent

We have already discussed the indications of integrability associated
to the super-NLS equation arising from its bihamiltonian structure.
Moreover we are aware of the results of $\cite{das}$ concerning the
integrability. In this section we will show that the equation
(\ref{superNLS}) deserves the name of super-NLS
hierarchy by explicitly associating to it a reduction of the super-KP
operator.
Before doing that let us spend some words on the supersymmetric (with
graded derivative) version of the KP hierarchy. The standard
reference we follow in this case is \cite{manin}.\par
The super-KP operator is given by
\bea
L &=& D +\sum_{i=0}^\infty U_i (X) D^{-i}
\ena
where now $D$ is the fermionic derivative and the $U_i$'s are
superfields. For even values of $i$ they are fermionic, for odd
values bosonic.
In the following we will be interested only to the flows associated
to even (bosonic) time. For a discussion concerning odd-time flows
see e.g. \cite{ramos}.
The even-time flows are defined through
\bea
{\partial L \over \partial t_k} &=& [ {L^{2k}}_+, L]
\ena
where ${L^r}_+$ denotes the purely differential part of $L^r$. The
above flows
provide a set of equations for the infinite series of superfields
$U_i$. To derive such equations we recall that $D^{-1} = D
\partial^{-1}$ and the commutation rule (\ref{comrul}) can be
employed.\par
If we set the constraint
\bea
DU_0 + 2 U_1 &=& 0
\ena
then ${L^2}_+ = D^2=\partial$ and the first flow is trivial. With the
above constraint we get
\bea
{L^4}_+ &=& D^4 +FD +B
\ena
where
\bea
F &=& 2 DU_1\nonumber\\
B &=& 4U_3 + 2 DU_2 -6 U_1U_1
\ena
The second flow ($k=2$) is non-trivial and provides the following set
of equations
\bea
{\partial {U_{2n}}\over \partial t_2} &=& {U_{2n}}^{(4)} + 2
{U_{2n+2}}^{(2)} +F {U_{2n}}^{(1)} + 2 F
U_{2n+1}-U_{2n-1}B^{(1)}+\nonumber\\
&& \sum_{r=1}^{n-1} (-1)^{r+1}
\left( \begin{array}{c} n-1\\ r \end{array}\right)
(U_{2n-2r} B^{(2r)} +U_{2n-2r-1}B^{(2r+1)})+\nonumber\\
&&  \sum_{r=1}^n (-1)^r
\left( \begin{array}{c} n\\ r \end{array}\right)
U_{2n-2r+1} F^{(2r)}
\ena
for the fermionic superfields, and
\bea
{\partial {U_{2n-1}}\over \partial t_2} &=& {U_{2n-1}}^{(4)} +
2{U_{2n+1}}^{(2)} +F{U_{2n-1}}^{(1)} - F^{(1)} U_{2n-1} +\nonumber\\
&& \sum_{r=1}^{n-1} (-1)^{r+1}
\left( \begin{array}{c} n-1\\ r \end{array}\right)
( U_{2n-2r-1}B^{(2r)} + U_{2n-2r-1}F^{(2r+1)}
+U_{2n-2r}F^{(2r)})\nonumber\\
\ena
for the bosonic ones.\par
In order to define the reduced super-KP operator we compare this
flows with the set of equations
\bea
{\dot V_n} &=& \{ V_n, H_2\}
\ena
for the superfields $V_n = \Psi_- {\cal D}^n \Psi_+ $ introduced in
(\ref{superinv}), provided by the hamiltonian $H_2$ given in
(\ref{superhami}), with respect to the
(\ref{zerosusyalg},\ref{susyalg}) Poisson brackets structure.\par
We get the following equations, for respectively fermionic and
bosonic superfields
\bea
{\partial V_{2n+1} \over \partial t_2 } &=& \partial^2 V_{2n+1} -
2\partial V_{2n+3} - V_{2n+1}\partial V_0 -V_{2n}\partial V_1
+\nonumber\\
&& \sum_{k=0}^{n-1}
\left( \begin{array}{c} n\\ k \end{array}\right)
(V_{2k+1}  \partial^{n-k} DV_1 -V_{2k} \partial^{n-k+1}V_1)
\ena
and
\bea
{\partial V_{2n} \over \partial t_2 } &=& \partial^2 V_{2n} -
2\partial V_{2n+2}  -V_{2n}\partial V_0 +\nonumber\\
&& \sum_{k=0}^{n-1}
\left( \begin{array}{c} n\\ k \end{array}\right)
V_{2k}  \partial^{n-k} DV_1
\ena
In order to produce a consistent super-KP reduction we must be able
to fit the above equations in the corresponding equations for the
$U_i$ superfields. This can not be done, or at least we were unable
to do that, for the whole set of $V_n$ superfields. However the
following considerations can be made: we remark that
the equations of motion for bosonic superfields (labelled by an even
integer)
involve on the right hand side bosonic superfields only. It is
therefore
consistent with the dynamics to set all the bosonic superfields
$V_{2n}\equiv 0$.
We argue that this constraint should be imposed in order to proceed
to the right
supersymmetrization of the NLS hierarchy: indeed the corresponding
generators of the coset algebra in the bosonic case are given by the
$J_-{\cal D}^nJ_+$ fields, which implies having a single bosonic
field for each integral value of the spin
($n+2$). In the supersymmetric theory one expects that
the fermionic counterparts should be associated to such fields: for
each half-integral value of the spin one should have a single
fermionic field. The set of superfields $V_n$, $n=0,1,2,...$ is in
this respect highly redundant: it provides two bosons and two
fermions respectively for each integer and half-integer spin value
$s\geq {\textstyle{3\over 2}}$, plus a single spin $1$ bosonic field
arising from $V_0$ which plays no role in the NLS hierachy.
To get rid of this redundancy, a constraint which kills the extra
degrees of freedom should be imposed. A constraint which allows doing
that is just provided by setting
\bea
V_{2n} &=& 0 \quad\quad for \quad n=0,1,2,..
\label{boscon}
\ena
It is remarkable the consistency of this constraint with the
dynamics, as we
have just pointed out.\par
After taking into account of (\ref{boscon}), the equation for the
fermionic superfields $V_{2n-1}$ is reduced to
\bea
{\dot V}_{2n-1} &=&
\partial^2 V_{2n-1} - 2\partial V_{2n+1} +\sum_{k=0}^{n-1}
\left( \begin{array}{c} n-1\\ k \end{array}\right)
V_{2k-1}  \partial^{n-k} DV_1
\ena
It is immediately checked at this point that a consistent reduction
of the super-KP hierarchy is recovered by setting
\bea
U_{2n-1} &=& 0 \nonumber\\
U_{2n} &=& {\textstyle{1\over 2}} (-1)^n V_{2n-1}\quad\quad for\quad
n=1,2,...
\ena
The corresponding reduced super-KP operator can be compactely written
as
\bea
L &=& D +{\textstyle{1\over 2}} \Psi_- {\cal D}^{-2} {\Psi_+}^{(1)}
\ena
with ${\Psi_+}^{(1)}={\cal D} \Psi_+$.\par
The integrability properties of the super-NLS hierarchy are
established due to the existence of such Lax operator.

\section{The $N=2$ formalism.}

\indent

Let us introduce here the framework and conventions for working in a
manifestly
supersymmetric $N=2$ formalism.\par
The $N=2$ superspace is parametrized by the bosonic $x$ coordinate
and two
grassmann variables $\theta, {\overline\theta}$. A generic superfield
is then expanded as
\bea
\Phi(X) &=& \phi (x) +\theta f(x) +{\overline \theta} {\overline
f}(x) + \theta{\overline\theta} g(x)
\label{n2super}
\ena
The $N=1$ case is recovered when letting
$\theta={\overline\theta}$.\par
Two spinor derivatives $ {\tilde D}, {\overline D} $ are defined as
\bea
{\tilde D} &=& {\partial \over \partial\theta} +{\overline \theta }
\partial_x\nonumber\\
{\overline D} &=&{\partial \over \partial{\overline\theta}} +{ \theta
} \partial_x
\ena
They satisfy the relations
\bea
{\tilde D}^2 = {\overline D}^2 &=& 0\nonumber\\
\{ {\tilde D}, {\overline D} \} &=& 2\partial_x
\ena
It is convenient (we come later on this point) to describe the
$N=2$ theory in terms of constrained superfields, namely the chiral
(${\tilde \Psi}$) and antichiral (${\overline \Psi}$) superfields,
defined respectively by
\bea
{\overline D} {\tilde \Psi} &=& 0
\nonumber\\
{\tilde D} {\overline \Psi} &=& 0
\ena
Due to the above relation the derivated superfields
${\tilde D}\Phi $ and ${\overline D}\Phi$ are respectively antichiral
and chiral superfields.\par
The condition of chirality implies the following expansions in
component fields
\bea
{\tilde A} &=& a(x) + \theta \alpha (x) + \theta{\overline \theta}
a(x)'\nonumber\\
{\overline B} &=& b(x) +{\overline \theta} {\beta}(x)
-\theta{\overline\theta} b(x)'
\ena
and the derivated superfields are
\bea
{\tilde D} {\tilde A} &=& \alpha (x) + 2{\overline\theta} a(x)'
-\theta{\overline\theta} a(x)''\nonumber\\
{\overline D} {\overline B} &=& \beta (x) + 2{\theta} b(x)'
+\theta{\overline\theta} b(x)''
\ena
It is remarkable that chiral and antichiral superfields
can be expressed as $N=1$ superfields in relation with the
superspaces
\bea
{\hat X} &=& ({\hat x} = x+\theta{\overline \theta},
\theta)\nonumber\\
{\check X} &=& ({\check x} = x-\theta{\overline \theta},{\overline
\theta})
\ena
respectively.\par
Moreover if we introduce the $N=1$ spinor derivative $D$ as
\bea
D&=& D_X ={\partial\over\partial\theta} + 2\theta {\partial\over
\partial x}
\label{newder}
\ena
(allowing a factor $2$ difference with respect to the convention used
in the previous sections), then we can write the derivated
superfields as
\bea
{\tilde D} {\tilde A} &\equiv& D {\tilde A}|_{\check X}\nonumber\\
{\overline D} {\overline B} &\equiv& D {\overline B}|_{\hat X}
\ena
The existence of the $N=1$ superfield representation for chiral and
antichiral
superfields is particularly useful for our purposes because it allows
defining the $N=2$ supersymmetric theory in terms of the $N=1$
superfield formalism developped in the previous sections. In
particular we can define
super-Poisson brackets structures as done before: they will depend on
the $N=1$ supersymmetric
delta-function already encountered (and the (\ref{newder}) derivative
acting on it).\par
The supersymmetric line integral for chiral and antichiral
superfields are given respectively by
\bea
d{\hat X} &\equiv& d X\quad\quad\quad X=(x,\theta)\nonumber\\
d{\check X} &\equiv&d{\overline X}\quad\quad\quad{\overline X}
=(x,{\overline\theta})
\ena
The two equivalence relations are due to the fact that the term
proportional to ${\theta{\overline\theta}}$ is a total derivative for
both chiral and antichiral superfields .\par
Let us spend now some more words about using (anti)-chiral
superfields to describe $N=2$ theories. The dynamics of a real
superfield $\Psi $ can always be
recovered from the dynamics of two conjugated
chiral and antichiral superfields
$ {\tilde \Psi}$, $ {\overline \Psi}$ (we recall that the $g(x)$
field appearing in (\ref{n2super}) is just an auxiliary field,
dynamically determined in terms of
the component fields $\phi , f, {\overline f}$).\par
It turns out that the $N=2$ dynamics can be expressed by using two
conjugated sets of equations of motion for chiral and antichiral
superfields. Such equations of motion are defined in terms of
conjugated (anti-)chiral hamiltonians whose combination gives a
single real hamiltonian. For integrable systems the
dynamics can also be expressed through two chirally conjugated Lax
operators
whose combination provide a single real Lax operator. Further details
concerning
such construction can be found in (\cite{IT}). In the following we
will define
the $N=2$ super-NLS equation in terms of these two conjugated sets of
chiral
superfields.

\section{The $N=2$ super-NLS hierarchy.}

\indent

Let us introduce now the $N=2$ super-NLS hierarchy, extending to this
case
the procedure already worked out for the bosonic and $N=1$ NLS
theories. \par
According to the discussion developped in the previous section, it is
clear
that now we should define our $N=2$ hierarchy by ``doubling" the
number of superfields
of the $N=1$ case: we should look for two (chiral and antichiral)
covariant derivatives defined in terms of the spin
${\textstyle{1\over 2}}$ superfields ${\tilde \Psi_0}, {\overline
\Psi_0}$. Moreover we should have two
sets of opposite charged (anti-)chiral superfields
${\tilde\Psi}_\pm$, ${\overline\Psi}_\pm$. These two sets of
superfields should be seen as chirally conjugated. \par
Let us define now the two conjugated covariant derivatives: we
introduce first the conjugate spin ${\textstyle{1\over 2}}$
superfields ${\tilde\Psi}_0, {\overline\Psi}_0$.  There is a freedom
in choosing the normalization
condition for their super-Poisson brackets algebra. Let us fix it by
assuming
\bea
\{ {\tilde\Psi}_0(X),{\tilde\Psi}_0(Y)\}&=& \{
{\overline\Psi}_0(X),{\overline\Psi}_0(Y)\} = D_Y\Delta
(X,Y)\nonumber\\
 \{{\tilde\Psi}_0(X),{\overline\Psi}_0(Y)\}&=&
0
\label{superu1}
\ena
with $D_Y$ given by (\ref{newder}).\par
Next the notion of covariant superfield can be introduced: $V$ is
said covariant with charges $({\tilde q},{\overline q})$ if it
satisfies the relations
\bea
\{ {\tilde\Psi_0}(X),{V(Y)}\}&=& {\tilde q}\Delta (X,Y) V(Y)
\ena
and an analogous one with ${\tilde \cdot}\mapsto {\overline\cdot}$
.\par
The covariant derivative ${\cal D}$, mapping covariant superfields of
charges $({\tilde q},{\overline q})$ into superfields of the same
charge, is in this case
given by
\bea
{\cal D} V &=& (D +{\tilde q}{\tilde \Psi_0} +{\overline q}{\overline
\Psi_0}) V
\ena
At this point we have all the ingredients to define the complete
supercurrents
algebra involving
${\tilde\Psi}_0,{\tilde\Psi}_\pm,{\overline\Psi}_0,{\overline\Psi}_\pm$
which allows us to define the $N=2$ super-NLS theory. After a little
inspection
one can realize that our game can be played by simply postulating
such
algebra as given by two separated copies of the $N=1$
(\ref{zerosusyalg},\ref{susyalg}) algebra.
A fundamental point is that now, in order to recover the non-trivial
equations of motion which involve together chiral and antichiral
superfields, the two
$N=1$ supercurrents algebras should mix chiral and antichiral
superfields.\par
We can assume the two copies being given by (${\tilde\Psi}_- ,
{\overline\Psi}_0,{\overline\Psi}_+$)
and (${\overline\Psi}_-, {\tilde\Psi}_0, {\tilde\Psi}_+$), with the
following charges for the
${\tilde\Psi}_\pm,{\overline\Psi}_\pm$ superfields
\bea
{\tilde\Psi}_-&\equiv& (0,-1)\nonumber\\
{\overline\Psi}_+&\equiv& (0,1)\nonumber\\
{\overline\Psi}_-&\equiv& (-1,0)\nonumber\\
{\tilde\Psi}_+ &\equiv& (1,0)
\ena
The complete algebra is given by
\bea
\{{\overline\Psi}_0 (X), {\overline\Psi}_+(Y) \} &=& \Delta
(X,Y){\overline \Psi}_+(Y) \nonumber\\
\{{\overline\Psi}_0 (X), {\tilde\Psi}_-(Y) \} &=& -\Delta
(X,Y){\tilde \Psi}_-(Y) \nonumber\\
\{{\overline\Psi}_+ (X), {\tilde\Psi}_-(Y) \} &=&
(D_Y -{\overline\Psi}_0(Y) )
\Delta (X,Y) = {\cal D}_Y \Delta (X,Y)\nonumber\\
\{{\tilde\Psi}_0 (X), {\tilde\Psi}_+(Y) \} &=& \Delta (X,Y){\tilde
\Psi}_+(Y) \nonumber\\
\{{\tilde\Psi}_0 (X), {\overline\Psi}_-(Y) \} &=& -\Delta
(X,Y){\overline \Psi}_-(Y) \nonumber\\
\{{\tilde\Psi}_+ (X), {\overline\Psi}_-(Y) \} &=&
(D_Y -{\tilde\Psi}_0 (Y) )
\Delta (X,Y)= {\cal D}_Y \Delta (X,Y)
\label{susyalg2}
\ena
Together with (\ref{superu1}).
All other super-Poisson brackets are vanishing.\par
There exists of course a superWakimoto representation, provided by
two sets of
chirally conjugated superfields:
 the bosonic superfields
${\hat C}, {\check C}$ of weight $0$, and the fermionic ones ${\hat
B}, {\check B}, {\hat N}, {\check N} $ of weight ${\textstyle{1\over
2}}$. The $B$'s and $C$'s
superfields generate two coupled systems. \par
The superalgebra of the free Wakimoto
superfields is just provided by
\bea
\{ {\hat B}(X),{\hat C}(Y)\}&=&\Delta (X,Y)\nonumber\\
\{ {\hat C}(X),{\hat B}(Y)\}&=&\Delta (X,Y)\nonumber\\
\{ {\hat N}(X),{\hat N}(Y)\}&=&D_Y\Delta (X,Y)
\ena
and an equivalent relation with ${\hat\cdot}\mapsto{\check\cdot}$.
The superfields identifications are the same as in (\ref{superwak}):
\bea
{\overline \Psi}_+ &=& {\hat B}\nonumber\\
{\overline \Psi}_0 &=& -{\hat B}{\hat C} + {\hat B}\nonumber\\
{\tilde \Psi}_- &=& -{\textstyle{1\over 2}} {\hat B}{\hat C}^2 +{\hat
C}{\hat N}- D{\hat C}
\ena
and the analogous relations involving the second set of
superfields.\par
Inspired by the $N=1$ results we can define at this point our
dynamics
as determined by the two conjugated sets of (anti-)chiral
hamiltonians in involution. The first two (${\tilde H}_{1,2}$ and the
conjugates ${\overline H}_{1,2}$) are given by
\bea
{\tilde H}_1 &=& \int d{ X}{\tilde{\cal H}}_1 =\int dX ({\tilde
\Psi}_- {{\cal D}}{\overline \Psi}_+)\nonumber\\
{\tilde H}_2 &=& \int dX {\tilde {\cal H}}_2 =\int d{ X} ({\tilde
\Psi}_-
{{\cal D}}^3{\overline \Psi}_+)
\ena
and
\bea
{\overline H}_1 &=& \int d{\overline X}{\overline {\cal H}}_1 =\int
d{\overline X}( {\overline \Psi}_- {{\cal D}}{\tilde
\Psi}_+)\nonumber\\
{\overline H}_2 &=&  \int d{\overline X}{\overline {\cal H}}_2 =\int
d{\overline X} ({\overline \Psi}_-
{{\cal D}}^3{\tilde \Psi}_+)
\ena
The real hamiltonians are given by
\bea
H_{1,2} &=& {\tilde H}_{1,2} + {\overline H}_{1,2}\nonumber\\
\ena
They are invariant under the $N=2$ supersymmetry transformations
\bea
\delta {\tilde\Psi}_\pm &=& \pm\varepsilon {{\cal D}}{\overline
\Psi}_\pm\nonumber\\
\delta {\overline\Psi}_\pm &=& \pm{\overline \varepsilon} {{\cal
D}}{\tilde \Psi}_\pm
\ena
Moreover the hamiltonian densities ${\tilde {\cal H}}_j,
{\overline{\cal H}}_j$
have by construction vanishing Poisson brackets with respect to the
subalgebra
generators ${\tilde\Psi}_0,{\overline\Psi}_0$, namely they are in the
commutant.
\par
The equations of motion are introduced through the following
equations
\bea
{\partial\over\partial t_j } { F} &=& \{ H_j, F\}
\ena
After using the algebraic relations (\ref{superu1},\ref{susyalg2}),
and taking into account that we can consistently set
\bea
{\tilde\Psi}_0={\overline\Psi}_0 &=& 0
\ena
we get the flows:
\bea
{\partial\over\partial t_1 } {\tilde\Psi}_\pm &=& {\tilde\Psi}_\pm
'\nonumber\\
{\partial\over\partial t_1 }{\overline\Psi}_\pm &=&
{\overline\Psi}_\pm '
\ena
and
\bea
{\partial\over\partial t_2} {\tilde\Psi}_\pm &=& \pm
{\tilde\Psi}_\pm'' \mp
{\tilde\Psi}_\pm {\overline D} ({\overline\Psi}_\mp {\tilde D}
{\tilde \Psi}_\pm )\nonumber\\
{\partial\over\partial t_2 }{\overline\Psi}_\pm &=& \pm
{\overline\Psi}_\pm'' \mp
{\overline\Psi}_\pm {\tilde D} ({\tilde\Psi}_\mp {\overline D}
{\overline \Psi}_\pm )
\ena
The second flow provides the two-components $N=2$ super-NLS
equation.\par
Notice that the chirality condition is respected by the equations of
motion as it should be.\par
On the right hand side chiral and antichiral superfields are coupled
together in the non-linear term. This ensures the theory having the
genuine feature of a non-trivial
$N=2$ supersymmetry.
The $N=1$ equation is recovered
by assuming $\theta={\overline\theta}$ which implies
${\tilde\Psi}_\pm ={\overline\Psi}_\pm$.\par
It is clear that one can straightforwardly repeat the same steps
as done in the $N=1$ construction. The same structures appear in this
case as well. Let us recall them briefly:\par
i) existence of a compatible bihamiltonian structure relating the
first two hamiltonians.\par
ii) $N=2$ generalization of the modified super-NLS equation arising
by the superWakimoto representation for the algebra
(\ref{superu1},\ref{susyalg2}).\par
iii) existence of the (coset) $N=2$ non-linear super-$\cw_\infty$
algebra, promoted to be a finite rational super-$\cw$ algebra. It is
linearly generated by the chargeless superfields
\bea
V_{2n} &=& {\tilde \Psi}_- {\cal D}^{2n}
{\overline {\Psi}}_+\nonumber\\
V_{2n+1} &=&{\tilde \Psi}_- {\cal D}^{2n+1}
{\overline {\Psi}}_+\nonumber\\
W_{2n} &=&  {\overline\Psi}_- {\cal D}^{2n}
{\tilde{\Psi}}_+\nonumber\\
W_{2n+1} &=&{\overline \Psi}_- {\cal D}^{2n+1}
{\tilde {\Psi}}_+
\ena
The fermionic superfields $V_{2n+1}$, $W_{2n+1}$ have half-integral
spin ${\textstyle {2n+3\over 2}}$. When evaluated at
${\tilde\Psi}_0={\overline\Psi}_0=0$ they are
respectively chiral and antichiral, and can be expressed as
\bea
V_{2n+1} &=&{\tilde \Psi}_-(2\partial )^n {\overline D}
{\overline {\Psi}}_+\nonumber\\
W_{2n+1} &=&{\overline \Psi}_- (2\partial )^n{\tilde D}
{\tilde {\Psi}}_+
\ena
 The bosonic superfields $V_{2n}$, $W_{2n}$ of spin $n+1$
have not a definite chirality. Notice that, as it should be, our
$N=2$ super-$\cw$ algebra admits a ``doubled" number of superfields
with respect to the $N=1$ case.\par
iv) existence of a dynamically consistent constraint, which allows
setting
the bosonic superfields $V_{2n}$, $W_{2n}$ equal to zero.
This implies in its turn a ``reduced dynamics" involving only the
chiral and antichiral fermionic
superfields; such a dynamics is particularly important because
it gives rise to a consistent reduction of the $N=2$
super-KP hierarchy provided by the two conjugate Lax operators of
definite chirality.
\par
These two conjugate Lax operators are given by
\bea
{\tilde L} &=& {{\cal D}} + {\tilde\Psi}_- {\cal D}^{-2}
{{\overline\Psi}_+}^{(1)}
\nonumber\\
{\overline L} &=& {{\cal D}} + {\overline\Psi}_- {\cal D}^{-2}
{{\tilde\Psi}_+}^{(1)}
\ena
where
\bea
{{\overline\Psi}_+}^{(1)}&=& {{\cal D}}{\overline\Psi}_+\nonumber\\
{{\tilde\Psi}_+}^{(1)}&=& {{\cal D}}{\tilde\Psi}_+
\ena
${\tilde L} $ is chiral, ${\overline L}$ antichiral.\par
Once expanded, they are expressed in terms of the $V_{2n+1}$,
$W_{2n+1}$ superfields respectively, which are invariants under the
$N=2$ Kac-Moody superalgebra (\ref{superu1}):
\bea
{\tilde L} &=& {\tilde{ D}} +\sum_{k=0}^\infty (-1)^kV_{ 2k+1}
\partial^{-k}\nonumber\\
{\overline L} &=& {\overline{ D}} +\sum_{k=0}^\infty (-1)^kW_{ 2k+1}
\partial^{-k}
\label{supkpred}
\ena
(we have replaced the covariant derivative with the standard one,
which is allowed when ${\tilde L}$, ${\overline L}$ act on chargeless
superfields).\par
The dynamics for the $V_{2n+1}$, $W_{2n+1}$ superfields derived in
terms of flows of the super-KP reduced operator (\ref{supkpred})
coincides with the
just mentioned ``reduced dynamics" of $V_{2n+1}$, $W_{2n+1}$ arising
from the hamiltonian formulation.
{}~\quad\\
\vfill
{\Large {\bf Conclusions}}

\indent

In this paper we have furnished a method to derive what we can call
(in analogy to the bosonic case) multi-superfields reductions of the
super-KP hierarchy, which are a further generalization of the
commonly studied generalized super-KdV hierarchies.\par
In the particular example here considered we obtained some new
results
concerning the form of the super-Lax operator, the connection with a
super$\cw_\infty$ algebra, the link with the modified super-NLS
equation, etc.\par
According to our ``coset method" the multifields reductions are
obtained from cosets of (in this case super) Kac-Moody algebras.\par
We would like to spend some words about the coset method and why it
deserves
being further investigated: it allows having a nice algebraic
interpretation
for the Poisson brackets structures of the theories involved; more
than that,
it could provide an algebraic classification of the multi-fields
(super) KP reductions if the attracting hypothesis that they are all
associated to cosets
proves to be correct. Since our method makes use of covariant
derivatives and is not based on a hamiltonian reduction (and
consequently on Dirac's brackets) it implies a nice free-fields
interpretation and mapping to modified hierarchies as explained in
the paper.
This could prove useful when discussing quantization (it is tempting
indeed to repeat our procedure for let's say the $q$-deformed affine
$sl(2)$ algebra).\par
In order to attack the most important point, concerning the
classification
of the (super) KP reductions some preliminary results will be needed:
we can mention for instance understanding the coset method
in the light of the AKS scheme, expliciting the connection between
the
(unreduced) KP hierarchy Poisson brackets structure and those coming
from the cosets, computing the associated $r$-matrices with methods
like those developped in \cite{rmat}. Such results are needed for a
formal proof of the statement that any coset gives rise to a
KP-reduction.
We will address all these points in forthcoming papers.
{}~\\~\\

\noindent
{\large{\bf Acknowledgements}}
{}~\\~\\
I wish to acknowledge useful discussions had
with L. Feher and P. Sorba.
{}~\\
{}~\\


\begin{thebibliography}{99}

\bibitem{GGKM} C.S. Gardner, J.M. Green, M.D. Kruskal and R.M. Miura,
Phys. Rev. Lett. 19 (1967), 1095.
\bibitem{mar} A. Marshakov, Int. Jou. Mod. Phys. A, Vol.8, n. 22
(1993), 3831.
\bibitem{DS} V. Drinfeld and V. Sokolov, Jou. Sov. Math 30 (1984),
1975.
\bibitem{ara} H. Aratyn, L.A. Ferreira, J.F. Gomes and A.H.
Zimerman,
Nucl. Phys. {B 402} (1993), 85; Preprint UIC-HEP-TH-93-05, hep-th
9304152.  H. Aratyn E. Nissimov and S. Pacheva, Phys. Lett. B 314
(1993) 41;
Preprint UICHEP-TH-94-2, hep-th 9401058.
\bibitem{bonora} L. Bonora and C.S. Xiong, Phys. Lett. B 285 (1992),
191;
Phys. Lett. {B 317}
(1993), 329.
\bibitem{bonora2} L. Bonora and C.S. Xiong, Preprint SISSA-171/93/EP,
BONN-HE-46/93, hep-th 9311070.
\bibitem{yuwu} F.~Yu and Y.S.~Wu, Phys. Rev. Lett. {68} (1992),
2996; Nucl. Phys. {B 373} (1992), 713.
\bibitem{bakas} I. Bakas and E. Kiritsis, Int. J. Mod. Phys. A7
(1992), 55;  J. Schiff, Preprint IASSNS-HEP-92-57, hep-th
9210029.
\bibitem{toppan} F. Toppan, Preprint ENSLAPP-L-448/93, hep-th
9312045, to appear in Phys. Lett. B.
\bibitem{feher}  L.~Feher, L.~ O'Raifeartaigh, P.~Ruelle and
I.~Tsutsui, Phys.
		Lett B283 (1992) 243.
\bibitem{DFSTR} F.~Delduc, L. Frappat, P. Sorba, F. Toppan
and E. Ragoucy,
		Phys. Lett. B. 318 (1993), 457.
\bibitem{feher2}  J. de Boer, L. Feher and A. Honecker, Preprint
ITP-SB-93-84,
hep-th 9312049, to appear in Nucl. Phys. B.
\bibitem{waki} M. Wakimoto, Comm. Math. Phys. 104 (1986), 605.
\bibitem{manin} Yu.I. Manin and A.O. Radul, Comm. Math. Phys. 98
(1985), 65.
\bibitem{others} P. Mathieu, Jou. Math. Phys. 29 (1988), 2499; W.
Oevel and
Z. Popowicz, Comm. Math. Phys. 139 (1991), 441; J.M.
Figueroa-O'Farrill, J. Mas and E. Ramos, Rev. Mod. Phys. 3 (1991),
479.
\bibitem{N2} T. Inami and I. Kanno, Int. Jou. Mod. Phys. A7 (Suppl.
1A) (1992), 419; S. Bellucci, E. Ivanov, S. Krivonos and A. Pichugin,
Phys. Lett. B 312 (1993), 463.
\bibitem{das} J.C. Brunelli and A. Das, Preprint UR-1344, hep-th
9403019.
\bibitem{roe} G.H.M. Roelofs and P.H.M. Kersten, Jou. Math. Phys. 33
(1992), 2185.
\bibitem{IT} E. Ivanov and F. Toppan, Phys. Lett. B 309 (1993), 289;
Mod. Phys. Lett. A9 (1994), 51.
\bibitem{dickey} L.A. Dickey, Soliton Equations and Hamilton Systems,
Adv. Series in Math. Phys. Vol. 12, World Sc. 1991.
\bibitem{polya} A. Polyakov, Int. J. Mod. Phys. {A 5} (1990),
833; M.~Bershadsky, Comm. Math. Phys. 139 (1991) 71.
\bibitem{fadtak} L.A. Faddeev and L.A. Takhtajan, Hamiltonian Methods
in the Theory of Solitons, Springer-Verlag 1987.
\bibitem{GMMOS} A. Gerasimov, A. Morozov, M. Olshanetskii, A.
Marshakov and S. Shatashvili, Int. Jou. Mod. Phys. A5 (1990), 2495.
\bibitem{DFRS} F. Delduc, L. Frappat, E. Ragoucy and P. Sorba,
Preprint ENSLAPP-AL-449/93, hep-th 9312041. To appear in Proceedings
of the Int. Conf. on Diff. Geom. Meth. in Theor. Phys., Ixtapa
(Mexico) 1993.
\bibitem{ramos} E. Ramos, Preprint QMW-PH-94-4, hep-th 9403043.
\bibitem{rmat} H. Aratyn, E. Nissimov, S. Pacheva and I. Vaysburd,
Phys. Lett. B 294 (1992), 167; B. Enriquez, S. Khoroshkin, A. Radul,
A. Rosly and V. Rubtsov, Preprint Ecole Polytechnique,Palaiseau, n.
1064 (1993).
\end{thebibliography}
\end{document}